\documentstyle[11pt]{article}
\newcommand{\al}{\alpha}

\newcommand{\ga}{\gamma}

\newcommand{\ep}{\epsilon}

\newcommand{\la}{\lambda}
\newcommand{\ka}{\kappa}

\newcommand{\si}{\sigma}

\newcommand{\De}{\Delta}

\newcommand{\be}{\begin{eqnarray}}
\newcommand{\ee}{\end{eqnarray}}
\newcommand{\jt}{\tilde{J}}

\newcommand{\lra}{\longrightarrow}

\newcommand{\ti}{\tilde}

\newcommand{\lan}{\langle}
\newcommand{\ran}{\rangle}

\newcommand{\rar}{\rightarrow}

\newcommand{\np}{\newpage}

\newcommand{\vs}{\vspace}
\newcommand{\nl}{\newline}
\newcommand{\nn}{\nonumber}

\newcommand{\hh}{\hat{h}}

\begin{document}


\thispagestyle{empty}

\vs*{-15mm}
\begin{flushright}
G\"oteborg ITP 94-26 \\
December 1994 \\
hep-th/9501084
\vs{8mm}\\
\end{flushright}

\begin{center}

{\huge{Construction of BRST invariant states\\
\vspace{3mm}
in $G/H$ WZNW models}} \\
\vspace{10 mm}
{\Large{Stephen Hwang}\footnote{tfesh@fy.chalmers.se}} \vspace{2mm}\\
{\Large{and}}
\vspace{2mm}\\
{\Large{Henric Rhedin}\footnote{hr@fy.chalmers.se}} \\
\vspace{4 mm}

Institute of Theoretical Physics \\
Chalmers University of Technology \\
and \\
G\"oteborg University \\

\vs{10mm}

{\bf{Abstract}} \end{center}
\begin{quotation}

\noindent We study the cohomology arising in the BRST formulation
of G/H gauged WZNW models, i.e. in which the states of the gauged
theory are projected out from the ungauged one by means of a BRST
condition. We will derive for a general simple group $H$ with arbitrary
level, conditions for which the cohomology is non-trivial. We show,
by introducing a small perturbation due to Jantzen, in the highest
weights of the
representations, how states in the cohomology, "singlet pairs",
arise from unphysical states, "Kugo-Ojima quartets", as the perturbation is
set to zero. This will enable us to identify and construct
states in the cohomology. The ghost numbers that will occur are $\pm p$,
with $p$ uniquely determined by the representations of the algebras involved.
Our construction is given in terms of the current modes and relies on the
explicit form of highest weight null-states given by
Malikov, Feigen and Fuchs.  \end{quotation}

\np
\setcounter{page}{1}
WZNW models are of great importance in the study of
conformal field theories.
In particular, the gauging of these models is an essential
part to be able describe
most conformal field theories. This gauging will lead to an
effective action, which is
BRST invariant \cite{KS} and apart from the ungauged WZNW models
based on the Lie group $G$ it contains an auxiliary and non-unitary
WZNW model based on the gauged subgroup $H$ of $G$,
and a ghost sector.  In \cite{HR} we investigated the implications of
the BRST symmetry,
when applied in an operator formulation of these models. There it was
shown that under
certain restrictions, the BRST approach gave a consistent operator
formulation of the model
and the only states that survived were ghost free and satisfied
the usual highest weight
conditions w.r.t. $\hh$, the affine Lie algebra corresponding
to $H$.  The restrictions
were on the choice of representations: integral representations of the
ungauged WZNW model and a corresponding selection of
representations of the
auxiliary WZNW model.

In this and a subsequent paper \cite{Hw}, we will study a much more
general set of
representations, namely arbitrary highest weight representations
for both the original
and auxiliary sectors. The motivation for this is that the states
in the cohomology for
non-zero ghost numbers, which we will show are present for more
general representations,
may be of importance in the determination of the complete
physical spectrum in a specific
model. In particular, it is generally believed to be true
for topological theories.

 We will here concentrate on the construction of the states in
the cohomology. In \cite{Hw} a more detailed analysis of the
cohomology will be presented. Some of the results derived
there will be needed in the present work, although most of
the analysis presented here will be self-contained.  Our
analysis will always be confined to the relative cohomology
i.e. in which the $b$-ghosts corresponding to the Cartan generators are
required to annihilate states. We will derive general
conditions and  consistency equations for the non-trivial
states in the cohomology and then give a procedure for constructing
such states.

Our construction relies basically on the results and
techniques presented by us previously \cite{HR} and
the construction of highest weight null-states given
by Malikov, Feigen and Fuchs \cite{MFF}. Crucial is
also a trick due to Jantzen \cite{Ja} to perturb
representations. With this perturbation one will
clearly see how the non-trivial states in the
cohomology, {\it "singlet pairs"}, emerge from
the set of trivial and non-invariant states,
{\it "quartets"} in the terminology of Kugo and
Ojima \cite{KO}. The ghost numbers $\pm q$ of the
singlet pairs, will for fixed representations of the
original and auxiliary sectors, have one fixed value
of $q\neq 0$. These values will be the only ones possible
\cite{Hw}. We have, however, been unable to determine the
dimensionality of the
cohomology. In the simple examples that we have checked, the
dimensionality is one.

Our analysis will be for $G/H$ gaugings with a general simple
group $H$ of arbitrary level. However, there are some limitations
due to the lack of complete decomposability of representations
of $\hat{g}$ w.r.t.
$\hat{h}$. In two cases it is known that such limitations do not
exist, namely if we consider
integrable representations of $\hat{g}$ or if $G=H$. Thus our work
will be completely general for these cases. In the generic case,
however, one does not have complete decomposability. Then our
analysis must be interpreted with some care. The implications of
the lack of decomposability is that certain null-states of
$\hat{h}$ will not be null-states w.r.t. $\hat{g}$. This in turn
implies that states in the cohomology of $H/H$ models, which are
based on these null-states of $\hat{h}$ will no longer be in the
cohomology in the $G/H$ case.
Since the presence of states at ghost numbers $\pm q$
will imply states of ghost numbers $\pm (q+1),\pm (q+2),\ldots$
(for other representations), infinitely many states may drop out.

In \cite{BMP2}-\cite{HY}
some investigations of the cohomology using free field
techniques were presented. In particular, in \cite{BMP2},
conjecturing the cohomology of the Felder operator to be
one-dimensional for exaclty one degree it was shown, quite
generally, that
this implied an analogous result for the BRST cohomology in the
case of integrable representations,
i.e. one-dimensional at exactly one ghost number and
its negative. The results presented here and in \cite{Hw}
will confirm this result as far as the ghost numbers are
concerned, but as stated above, we have no results concerning the
dimensionality.

We start by giving a qualitative explanation of this note. One of the
basic ingrediences is a theorem due to Kugo and Ojima \cite{KO}, which
by using the BRST symmetry classifies the states of an irreducible
state-space with a well-defined and finite inner product. The theorem
states that it is possible to choose a basis consisting of states of
the three categories: \vs{2mm} \\
\noindent {\it singlets}: BRST non-trivial states $|S_0\ran$
at ghost number zero, $\lan S_0|S_0\ran\neq0$ \vs{1mm}\\
\noindent {\it singlet pairs}: BRST non-trivial states $|S_q\ran,
|\bar{S}_{-q}\ran$ of opposite ghost number, \nl $\lan\bar{S}
_{-q}|S_q\ran\neq0$\vs{1mm}\\
\noindent{\it quartets}: BRST trivial states and their conjugates
represented
 by the states \nl $|S_{q-1}\ran,\ |S_q\ran,\ |\bar{S}_{-q}\ran,\
|\bar{S}_{-q+1}\ran$, where the index denotes the ghost number, and \nl
$\lan \bar{S}_{-q}|S_q\ran\neq0$, $\lan \bar{S}_{-q+1}|S_{q-1}\ran\neq0$,
$\hat{Q}|S_{q-1}\ran=|S_q\ran$,
and $\hat{Q}|\bar{S}_{-q}\ran=|\bar{S}_{-q+1}\ran$. \vs{2mm}

\noindent Here $\hat{Q}$ is the appropriate BRST operator.

The state space in our case consists of highest weight Verma modules
over affine Lie algebras. If the Verma module is irreducible then
one may show
that only singlets and quartets appear \cite{HR}. For general weights
on the weight
lattice the Verma module is reducible, and singlet pairs will also
be present. The
basic idea behind our construction is to use a trick due to Jantzen \cite{Ja},
in which
one adds a small perturbation to the highest weight $\la\rar\la+\ep z$ where
$z$ is an appropriate vector. For $0<\ep\ll 1$ the Verma
module is irreducible, so we will again only have singlets and quartets.
In the limit $\ep\rar0$ the quartets $|S_{q-1}\ran_{\ep},\ |S_q\ran_{\ep},\
|\bar{S}_{-q}\ran_{\ep}\ |\bar{S}_{-q+1}\ran_{\ep}$ at non-zero ghost number
may fall into one of the following categories: i) the quartet becomes an
ordinary quartet of the reducible Verma module, ii) all four states in
the quartet
become null-states, iii) two states of the quartet turn into
null-states and the remaining two become BRST non-trivial. For the
third case,
which is the interesting one, we find that the latter states form a
singlet pair and
these states are non-trivial in the cohomology. The idea is to
identify the quartets
which in the limit $\ep\rar0$ give singlet pairs, in particular
to identify the states in
the quartets that will become appropriate null-states. This will be one of
 the key results in our work. In particular, we will show that the
states $|S_q\ran_{\ep}$ of the quartet, in the relevant case, will
have positive
 ghost numbers, while $|\bar{S}_{-q}\ran_{\ep}$ will have negative
ghost numbers i.e. that $\hat{Q}|S_{q-1}\ran_{\ep}=\epsilon
|S_q\ran_{\ep}$ and $\hat{Q}|\bar{S}_{-q}
\ran=|\bar{S}_{-q+1}\ran$ where $lim_{\ep\rar0}
|S_{q-1}\ran_{\ep}=|N_{q-1}\ran$ and $lim_{\ep\rar0}|\bar{S}
_{-q+1}\ran_{\ep}=|\bar{N}_{-q+1}\ran$ are
 null-states.
We will also find that it is sufficient for our construction to know the
highest weight null-states
 of the Verma module, and these are given by Malikov, Feigin
and Fuchs \cite{MFF}.  \\

We consider a Lie group $G$ valued WZNW model gauged with an anomaly
free subgroup $H$. The resulting action is decomposable into three
parts \cite{GK},\cite{KPSY},
the original WZNW action, an auxiliary $H$ valued WZNW action,
and a Fadeev-Popov ghost part. The different sectors have
symmetry algebras of
affine Lie type $\hat{g}_k$, $\hat{h}_{\ti k}$, and $\hat{h}_{k_{gh}}$,
respectively, of levels  $k$, $\ti{k}=-k-2c_h$, and $k_{gh}=2c_h$ where
$c_h$ is the quadratic Casimir of the adjoint representation of the
Lie algebra $h$. We take $k>-c_h$
since the case $k<-c_h$ may be considered by interchanging the
original and
auxiliary sectors. It is assumed that the state space
decomposes into a $\hat{h}_k$ highest weight Verma module
$M^h_{\la}$, an auxiliary $\hat{h}_{\ti{k}}$ highest weight
Verma module $\widetilde{M}^h_{\ti{\la}}$
and a ghost module ${\cal F}$. Define a
BRST operator on the space $M^h_{\la}\times\widetilde{M}^h_{\ti{\la}}
\times{\cal F}\equiv M^h_{\la,\ti{\la}}\times{\cal F}$, which is
of the form \cite{KS}
\be
Q=\oint\frac{dz}{2i\pi}\left[:c_a(z)(J^a(z)+\jt^a(z)):-
\frac{i}{2}f^{ad}_{\ \ e}
:c_a(z)c_d(z)b^e(z):\right]\label{brstq},
\ee
where $c_a(z)$ and $b^a(z)$ are ghost fields.

We will here study the relative cohomology  ${\rm \hat{H}}
^p(\hat{h}_k,\hat{h}_{\ti{k}};\ldots)$
which is the cohomology of the sub-complex
defined such that all states on the subspace satisfy $b_0^i
|\phi\ran\!=0$, $i=1,...,r_h$,
where $r_h$ is the rank of $h$. It is then appropriate to
consider the following  decomposition of the BRST charge
$Q=\hat{Q}+c_{0,i}J^{i,tot}_0+M_ib_0^i$.
Here $J^{i,tot}_0=\{Q,b^i_0\}$ is the Cartan generator of
the total current i.e. the sum of the currents of the original,
auxiliary and ghost sectors. On the subspace $\hat{Q}$ is nilpotent, so it
makes sense to use it as a BRST operator.

The Verma module is reducible if and only if it contains highest weight null
vectors
that are not proportional to the primary state $|0;\la\ran$. We define
the irreducible $\hat{h}$ module $L^h_{\la}$ as the $\hat{h}$ highest
weight Verma
module modulo the maximal proper Verma submodule the latter being the module
 containing all $\hat{h}$ Verma modules over highest weight
null-states not proportional to $|0;\la\ran$. States that are
in the maximal proper submodule are called null-states.
We define the module  $L^h_{\la,\ti{\la}}=L^h_{\la}\times
\ti{L}^h_{\ti{\la}}$. It is the relative cohomology on this
module which is our primary interest here.
We start, however, by considering the relative cohomology
of the full Verma module. \\
{\sc \underline{Theorem.} } {\it ${\rm \hat{H}}^p(\hat{h}_k,
\hat{h}_{\ti{k}};M^h_{\la,\ti{\la}})$ vanishes for $p<0$.}\\
{\sc \underline{Proof.} }The theorem follows from an analogous treatment
to the one presented in \cite{HR}, so we will content ourselves with
a brief sketch of the proof. We introduce a gradation of states
which essentially counts excitations: $\sharp \jt+\sharp b-\sharp c$.
The BRST equation may now be solved order by order in this
gradation. The BRST charge decomposes as $d_0+d_{-1}$, where
$d_0$ when acting on a state with maximal degree $N$, gives states
of maximal degree $N$ or lower. It is possible
to define a homotopy operator $\ka_0$ such that $(d_0\ka_0+\ka_0d_0)
|\psi\ran\propto|\psi\ran$ to highest order and for positive degrees.
Using this operation iteratively one eliminates higher orders in favour
of states at lower order. This may be continued until the maximal
degree is zero or lower.
{}From the gradation one finds that negative ghost number states will
always have positive degrees, and hence they must be BRST trivial.
This gives the theorem.  $\Box$ \\
\vs{-4mm}\\
{\sc \underline{Corollary 1.} } {\it Let $\widetilde{M}_{\ti{\la}}^h$
be irreducible then $\hat{\rm H}^p(\hat{h}_k,\hat{h}_{\ti{k}};
L^h_{\la}\times \widetilde{M}^h_{\ti{\la}})$ is zero for $p\neq0$, and
for $p=0$
contains states of the form $|0;\la\ran|0;\ti{\la}\ran|0\ran^+$.
$|0;\la\ran$ and $|0;\ti{\la}\ran$ are highest weight primaries w.r.t.
 $\hat{h}_k$ and  $\hat{h}_{\ti{k}}$, respectively. The weights satisfy
$\la+\ti{\la}+\rho=0$, where $\rho$ is the sum of the positive roots
of $h$. $|0\ran^+=\prod_{\al\in\De_h^+}c_0^{\al}|0\ran_{gh}$, where
$|0\ran_{gh}$
is the $Sl(2,R)$ invariant ghost vacuum and $\De^+_h$ is the set of
 positive roots of $h$. } \\
{\sc \underline{Proof.} } This case was the one considered in \cite{HR}.
The condition $\la+\ti{\la}+\rho=0$ comes from the requirement
$J^{tot,i}_0|\ldots \ran=0$.
$\Box$ \\
\vs{-4mm}\\
{\sc \underline{Corollary 2.} } {\it Let $|\bar{S}\ran\in{\rm
\hat{H}}^p(\hat{h}_k,\hat{h}_{\ti{k}};L^h_{\la,\ti{\la}})$ for $p<0$.
Then on $M^h_{\la,\ti{\la}}$, $\hat{Q}|\bar{S}\ran=|\bar{N}\ran$ for
some non-zero $|\bar{N}\ran\in M^h_{\la,\ti{\la}}/ L^h_{\la,\ti{\la}}$ }\\
{\sc \underline{Proof.} } Assume the contrary i.e. $\hat{Q}|\bar{S}\ran=0$.
Then by the theorem $|\bar{S}\ran=\hat{Q}|\bar{S}'\ran$. Now if $|\bar{S}
\ran\in L^h_{\la,\ti{\la}}$ then $|\bar{S}'\ran\in L^h_{\la,\ti{\la}}$.
This follows since $|\bar{S}'\ran$ cannot be a null-state because then
so would also $|\bar{S}\ran$. Hence $|\bar{S}\ran$ is BRST exact.
$\Box$ \\
\vs{-4mm}\\
We thus see, as stated in the introduction, that the non-trivial
BRST invariant states $|\bar{S}\ran$ of negative ghost number come from
the $\hat{Q}|\bar{S}\ran=|\bar{N}\ran$ part of the Kugo-Ojima
quartet. In order to see how the states evolve from the Kugo-Ojima
quartet we now introduce a trick due to Jantzen \cite{Ja}.
Let $z=\sum_iz^i\mu^i$ where the sum runs over all fundamental
weights $\mu^i$ and $z^i\neq0\ \forall i$. Take highest
weights $\la$, $\ti{\la}$, and introduce a perturbation
$\ep$ such that $\la\rar\la+\ep z$ and $\ti{\la}\rar\ti{\la}-\ep z$.
Define the module  $M^h_{\ep,\la,\ti{\la}}\equiv M^h_{\la+\ep z}
\times\ti{M}^h_{\ti{\la}-\ep z}$. For $0<\ep\ll 1$ it follows
from the Kac-Kazhdan determinant \cite{KK} that $M^h_{\ep,
\la,\ti{\la}}$ is irreducible. Then by corollary 1
$\hat{\rm H}^p(\hat{h}_k,\hat{h}_{\ti{k}};M^h_{\ep, \la,\ti{\la}})$
is zero for $p\neq0$, and only singlets and quartets
exists in the Verma module.
Consequently we see that if singlet pairs exist, they will
appear from quartets in the limit $\ep\rar 0$.

Using the perturbation $\ep$ we now look for quartets $|N\ran_{\ep},
|\bar{N}\ran_{\ep},|S\ran_{\ep},|\bar{S}\ran_{\ep}$,
with the properties: Two states $|N\ran_{\ep}$, $|\bar{N}\ran_{\ep}$
are null-states as $\ep\rar0$ and two are not null in this limit. In
addition, they couple as $_{\ep}\lan\bar{N}|N\ran_{\ep}\neq0$ for
$\ep\neq0$ and
$\lan\bar{S}|S\ran\neq0$. For later reference we will use a
terminology from Jantzen \cite{Ja} of filtered Verma modules.
We may decompose the Verma module as $M^h_{\ep,\la,\ti{\la}}=
M_{\ep}^{(0)}\supset M_{\ep}^{(1)}\supset ...$ where we define
the submodules $M^{(i)}_{\ep}$ to contain all states
$|S\ran_{\ep}$ such that $_{\ep}\lan\bar{S}|S\ran_{\ep}$ is
divisible by $\ep^i$. In the limit $\ep\rar 0$ this induces a
filtration of modules $M^h_{\la,\ti{\la}}=M^{(0)}\supset M^{(1)}\supset ...$ .
The states in $M^{(1)}/ M^{(2)}$ will be referred to as first generation
null-states. We may also note that the irreducible state space
$L^h_{\la,\ti{\la}}\equiv L^h_{\la}\times\tilde{L}^h_{\ti{\la}}$ is
isomorphic to $M^{(0)}/M^{(1)}$.

In order to explicitly construct the singlet pairs, we now study
which null-states will appear in the relevant quartets. For $p<0$
 we have from corollary 2 that $\hat{Q}|\bar{S}\ran=|\bar{N}\ran$
which means that $\hat{Q}|\bar{N}\ran=0$ on $M^h_{\la,\ti{\la}}$.
Now if there exists a state $|\bar{N}'\ran$ such that
$|\bar{N}\ran=\hat{Q}|\bar{N}'\ran$ then $|\bar{S}\ran$ is not in
the relative cohomology. This follows since $\hat{Q}(|\bar{S}\ran+
|\bar{N}'\ran)=0$ implies by corollary 2 that $|\bar{S}\ran+
|\bar{N}\ran$ is BRST exact. Consequently $|\bar{S}\ran$ is
BRST exact on $L^h_{\la,\ti{\la}}$, since here states are only
defined modulo null-states. We thus find that $|\bar{N}\ran$
must be in the relative cohomology on $M^{(1)}$. The
non-triviality of the relative cohomology at ghost number
$-q+1$ will, therefore, be essential for the non-triviality
of the relative cohomology at $p=-q$. Conversely, given a null-state
$|\bar{N}\ran$ of negative ghost number
which is in the relative cohomology on  $M^{(1)}/M^{(2)}$,
then by the theorem $|\bar{N}\ran=\hat{Q}|\bar{S}\ran$ for some
$|\bar{S}\ran$ which does not belong $M^{(1)}$. Hence, it must be in
the irreducible module $L^h_{\la,\ti{\la}}$ and thus the existence of
$|\bar{N}\ran\in M^{(1)}/M^{(2)}$ implies the existence of $|\bar{S}\ran$.
Furthermore,
knowing the former explicitely will, through the use of
the homotopy operator, make it possible to construct the latter.

For $p>0$ we have, for states $|N\ran_{\ep}$ and $|S\ran_{\ep}$
in a quartet for $\ep\neq0$, that $\hat{Q}|N\ran_{\ep}=
f(\ep)|S\ran_{\ep}$, and  $lim_{\ep\rar0}f(\ep)=0$ if $|S\ran$ is in
the cohomology.
Note that $\hat{Q}|S\ran_{\ep}=0$ for all values of $\ep$.
Since $\hat{Q}$ is linear in $J$ and $\jt$ we will in fact
have $f(\ep)\propto\ep$. We thus look for null-states
satisfying $\hat{Q}|N\ran=0$ for $\ep=0$. These can be found as follows.
Take an appropriate submodule $M_{\la',\ti{\la}}^h$.
Obviously all null-states of the form $|N\ran=\hat{Q}|N'\ran$
can be discarded. A possible choice is then to take $|N\ran$ to
belong to the cohomology $\hat{{\rm H}}^{p-1}(\hat{h}_k,\hat{h}_{\ti{k}};
L_{\la',\ti{\la}})$ where $L_{\la',\ti{\la}}$ is contained in
$M^{(1)}/M^{(2)}$. Then $\hat{Q}|N\ran=0$, and $\hat{Q}|N\ran_{\ep}=
\ep|S\ran_{\ep}\in M_{\ep,\la,\ti{\la}}^h$. A state $|S\ran$ found in
this way will then be a non-trivial state in the cohomology.
This may be proven by a more detailed analysis \cite{Hw}.
Thus non-trivial states at
ghost number $p$ may be obtained from non-trivial states at ghost number
$p-1$.
Another possibility is to
exchange the r$\hat{{\rm o}}$le of the modules $M^h_{\la}$ and
$\widetilde{M}^h_{\ti{\la}}$, but this will give us cohomologically
equivalent states to the ones above. This may be
proven using the homotopy operator introduced in the proof of the theorem.

Consider ghost number zero. We pick a Kugo-Ojima singlet
$|P_0\ran_{\ep}=|0,\la+\ep z\ran|0,\ti{\la}-\ep z\ran|0\ran^+$
with $\la+\ti{\la}+\rho=0$. The null-states of the quartet
are constructed by the substitution of the primary $|0,\la\ran$
by a corresponding highest weight null-state $|n_0\ran$ of first generation
i.e. $|P_0\ran_{\ep}\rar|N_0\ran_{\ep}$. $|S_1\ran$ is then found by
applying $\hat{Q}$,  $\hat{Q}|N_0\ran_{\ep}=\ep|S_1\ran_{\ep}$, and
taking the limit
$\ep\rar0$. It is clear that $|S_1\ran$ is of the form $|s_1\ran|0,
\ti{\la}\ran|gh\ran$ where $|gh\ran$ is the appropriate ghost state
which includes only $c$ ghosts. If we instead make the substitution
$|0,\ti{\la}
\ran_{\ep}\rar|\ti{n}\ran_{\ep}$,
then we would get a cohomologically equivalent state.

The next step is to find the state $|\bar{N}_0\ran$
which satisfies $\hat{Q}|\bar{N}_0\ran=0$ and couples to $|N_0\ran_{\ep}$
for $\ep\neq 0$. Since $_\ep\lan N_0|N_0\ran_\ep\sim\ep$ we can take
$|\bar{N}_0\ran=|N_0\ran+\ldots$\ . In order to find the rest
of $|\bar{N}_0\ran$, we use the following trick. Start by taking $|N_0\ran$.
Since it satisfies $\hat{Q}|N_0\ran=0$, we can use the homotopy
operator $\kappa_0$ introduced
below the theorem, but with the original sector
replacing the auxiliary sector.
Then we will get  $|N_0\ran=\hat{Q}|\bar{S}_{-1}\ran-|\widetilde{\bar{N}}_0
\ran$ for some
state $|\bar{S}_{-1}\ran$. We now take
$|\bar{N}_0\ran\equiv|N_0\ran+|\widetilde{\bar{N}}_0\ran$
and by this we have found the state $|\bar{S}_{-1}\ran$ at ghost
number minus one.

For ghost number two we proceed exactly as above. Let $|S_1\ran$ be a state
in the cohomology determined as above. Introduce Jantzen's perturbation,
and substitute the primary state with a highest weight null-state in the
original sector $|S_1\ran_{\ep}\rar|N_1\ran_{\ep}$. Act on this by the
BRST operator which will give $\hat{Q}|N_1\ran_{\ep}=\ep|S_2\ran_{\ep}$, etc.

Ghost number minus two follows essentially the same steps as minus one
although finding the BRST invariant null-state is more complicated.
We must, in the Verma module $M^h_{\la,\ti{\la}}$, find the appropriate
null-states at ghost number minus one, which belongs to the cohomology
at $p=-1$ of $M^{(1)}/M^{(2)}$. We take as a general ansatz
$\sum_ia_i|\bar{N}_{-1}\ran_i$ where $|\bar{N}_{-1}\ran_i$ is constructed
from states $|\bar{S}_{-1}\ran_i$ in the cohomology for an appropriate weight,
by substitution of a primary in $|\bar{S}_{-1}\ran_i$ with a
highest weight null-state in either the original or the
auxiliary sector. The requirement $\hat{Q}\sum_ia_i|\bar{N}_{-1}\ran_i=0$
determines the coefficients $a_i$. We know from the theorem of Kugo-Ojima,
that the number of solutions to this equation is exactly the same as the
number of solutions at ghost number two. Given a solution to $\hat{Q}
\sum_ia_i|\bar{N}_{-1}\ran_i=0$
we may iteratively use the homotopy operator $\ka_0$ to find a representative
in the cohomology at ghost number minus two.

In order to proceed to arbitrary ghost numbers we repeat this process
the desired number of steps following the schematic pattern depicted below.
\be
& &...\stackrel{\hat{Q}}{\lra}\ep|S_{p-1}\ran_{\ep}\lra|N_{p-1}
\ran_{\ep}\stackrel{\hat{Q}}{\lra}\ep|S_p\ran_{\ep}\lra|N_p
\ran_{\ep}\stackrel{\hat{Q}}{\lra}\ep|S_{p+1}\ran_{\ep}\lra...\nn \\
& &...\stackrel{\ka}{\lra}\hat{Q}|\bar{S}_{-p+1}\ran\lra|\bar{N}_{-p+1}
\ran\stackrel{\ka}{\lra}\hat{Q}|\bar{S}_{-p}\ran\lra |\bar{N}_{-p}\ran
\stackrel{\ka}{\lra}\hat{Q}|\bar{S}_{-p-1}\ran\lra... \nn
\ee
where $\ka$ is the extension of $\ka_0$ to all orders.

For the explicit construction we must know the explicit form of
highest weight null-states in the Verma module, which are given
in \cite{MFF}. Let us outline the construction in \cite{MFF} of highest
weight null-states. First we observe that the affine weight lattice for
$k>-c_h$
may be constructed from the set
of dominant weights and the Weyl group. Let $\hat{\mu}_0+\hat{\rho}/2$
be a dominant weight, i.e. $(\hat{\mu}_0+\hat{\rho}/2)\cdot\hat{\al}^i
\geq0$ for all simple affine roots $\hat{\al}^i$. $\hat{\rho}$ is defined
by $\hat{\rho}\cdot\hat{\al}^i=\hat{\al}^i\cdot\hat{\al}^i$. Take an affine
weight $\hat{\la}$. Then there exists a sequence of $\hat{\rho}$-centered
simple Weyl reflexions,  $\si_i^{\rho}(\mu_0)\equiv\si_i(\hat{\mu}_0+
\hat{\rho}/2)-\hat{\rho}/2$, such that $\si^{\rho}_{i_p}\si^{\rho}_{i_{p-1}}
...\si^{\rho}_{i_1}(\hat{\mu}_0)
<\si^{\rho}_{i_{p-1}}...\si^{\rho}_{i_1}(\hat{\mu}_0)$, and
$\hat{\la}=\si_{i_p}^{\rho}...\si_{i_1}^{\rho}(\hat{\mu}_0)$.
Furthermore, by requiring the "word" $(i_1,...,i_p)$ to be of minimum length,
$\hat{\mu}_0$ and this word are uniquely determined by $\hat{\la}$.
 Hence, we may use the set of Weyl reflexions and the dominant weights
to parametrize any weight $\hat{\la}$ for which $k>-c_h$. For the
auxiliary sector, which has $\ti{k}<-c_h$,
one may proceed similarly for $-\hat{\ti{\la}}-\hat{\rho}$.
For later reference we define in this parametrization, the length
$l_{\la}$ of a weight $\hat{\la}$, to be the number of entries in the word.
For the auxiliary sector the length $l_{\ti{\la}}$ of a weight
$\hat{\ti{\la}}$ is similarly defined using the weight
$-\hat{\ti{\la}}-\hat{\rho}$.
The $\hat{\rho}$-centered Weyl reflexion $\si^{\rho}_i$ of
$\hat{\mu}_0$ may be represented at the level of states as
$|\mu_0\ran\rar(f_i)^{\ga_i}|\mu_0\ran$, where $f_i$ is the affine
generator corresponding to $-\hat{\al}_i$, and $\ga_i$ is defined
from $\si^{\rho}_i(\hat{\mu}_0+\hat{\rho}/2)-\hat{\rho}/2-\hat{\mu}_0
\equiv-\ga_i\hat{\al}_i$. It is straightforward to verify that the
state $(f_i)^{\ga_i}|\mu_0\ran$ is a highest weight null-state.
The procedure may be repeated to produce two different highest
weight null-states $|\la_{i_1...i_n}\ran= (f_{i_n})^{\ga_{i_n}}
...(f_{i_1})^{\ga_{i_1}}|\mu_0\ran$ and $|\la_{j_1...j_m}\ran=
(f_{j_m})^{\ga_{j_m}}...(f_{j_1})^{\ga_{j_1}}|
\mu_0\ran$. By eliminating $|\mu_0\ran$ we may formally write
\be
|n\ran\equiv|\la_{i_1...i_n}\ran=(f_{i_n})^{\ga_{i_n}}...(f_{i_1})^
{\ga_{i_1}}(f_{j_1})^{-\ga_{j_1}}...(f_{j_{m}})^{-\ga_{j_{m}}}|
\la_{j_1...j_{m}}\ran
\ee
$|n\ran$ may not exist as a state in the Verma module over
$|\la_{j_1...j_m}\ran$ due to negative powers appearing on the right
hand side. It may, however, be shown \cite{MFF} that if the word of
$j_1...j_{m}$ may be obtained from $i_1...i_n$ by deleting $n-m$
letters of the latter word, then $|n\ran$ exists. The right hand
side may then be put in a well-defined form, where only positive
powers of the generators appear, and $|n\ran$ is the required
highest weight null-state. If $m=n-1$ then the null-states are
of first generation. We also note that if $|N\ran$ is a first
generation highest weight null-state of weight $\la',\ti{\la}'$
in $M_{\la,\ti{\la}}^h$ then $l_{\la'}+l_{\ti{\la}'}-l_{\la}-l_{\ti{\la}}=1$.

Let us now discuss what ghost numbers appear for which representations.
At ghost number zero we are constrained by $\la_0+\ti{\la}_0+\rho=0$,
which means that $l_{\la_0}-l_{\ti{\la}_0}=0$. We now recall that for
ghost number one we used the substitution $|0,\la_0\ran|0,\ti{\la}_0
\ran\rar|n_0\ran|0,\ti{\la}_0\ran$ where $|n_0\ran\in M^{(1)}/ M^{(2)}$.
This means that $l_{\la_1}-l_{\la_0}=1$ where $\la_1$ is the weight of
the state at ghost number one produced as above from $|n_0\ran_{\ep}|0,
\ti{\la}_0-\ep t\ran|0\ran^+$. We thus have that $l_{\la_1}-l_{\ti{\la}_0}=1$.
The argument may be repeated, and one realizes that for the generic
case the ghost numbers are given by $p=\pm|l_{\la}-l_{\ti{\la}}|$.
Here the absolute value arises from the fact that one may use the
substitution of primary to null-state in the auxiliary sector as well.
It is thus possible to fix the representation of say the original
sector and adjust the auxiliary sector to obtain any ghost number.
For the special case of integrable representations of $\hat{h}$ we
have $l_\la=0$ and consequently the ghost numbers are given by $p=\pm
l_{\ti{\la}}$.

We have this far performed our analysis on the module $L^h\times\tilde{L}^h$.
If we embed
$L^h$ into $L^g$ for a general
algebra $\hat{g}$, of which $\hat{h}$ is a subalgebra, our results change
in the following way. The theorem as well as the two
corollaries remain true. This follows since in proving these, we only used the
auxiliary sector for which state space is unchanged.
Thus it is still necessary for non-trivial states with negative
ghost number to satisfy
$\hat{Q}|\bar{S}\ran=|\bar{N}\ran$ for some null-state $|\bar{N}\ran$.
This null-state is always a sum of null-states in the auxiliary
as well as original sectors. Similarly for positive ghost numbers we have
the equation $\hat{Q}|N\ran_\ep=\ep|S\ran_\ep$. This shows that any
solution of the
H/H case will be contained in the G/H case, provided $|\bar{N}\ran$ is
also a null-vector w.r.t. $\hat{g}$.

Now, if the irreducible $\hat{g}$ module does not
completely decompose into a sum of irreducible $\hat{h}$ modules, then
this implies the existence of at least one highest weight null-state
w.r.t. $\hat{h}$
which is not a null-state w.r.t. $\hat{g}$ (see \cite{HR}). If this in turn
implies that $|\bar{N}\ran$  is not null, then $|\bar{S}\ran$
is not BRST invariant and consequently will no longer be in the cohomology.
Similarly for positive ghost
numbers, we have states
$|S\ran$ satisfying $\hat{Q}|N\ran_\epsilon=\epsilon |S\ran_\epsilon$.
This relation will still be true, even in the case when the
conjugate state to
$|N\ran$ is no longer a null-state. However, then $|S\ran$ must be BRST exact.
In our construction we have shown how non-trivial states at ghost
number $\pm p$
generate new non-trivial states at $\pm(p+1),\pm(p+2),\ldots$ .
If a particular
singlet pair will drop out of the cohomology, it may therefore imply that
many, possibly infinitely many, states will drop out of the cohomology.
If this is actually the case is still an open question.

For the special case of integrable representations
of the original sector, it has been proved \cite{KP} that the irreducible
$\hat{g}$ module decomposes completely into a sum of irreducible $\hat{h}$
modules.
Then it is easy to show that every null-state of $\hat{h}$ is also a
null-state of $\hat{g}$, from which it follows that the analysis and
construction given in this paper apply without restrictions. \\

Let us end by presenting some explicit examples for the choice of
$\hat{h}=\widehat{su}(2)$ with affine generators $J^+_n,J^-_n$ and
$J^3_n$ . We will content ourselves with giving a few states, and only for
ghost numbers $\pm 1,\pm 2$.

We start with the primary satisfying $j_0+\ti{j}_0+1=0$: $|j_0=-
\frac{k}{2}-2\ran|\ti{j}_0=
\frac{k}{2}+1\ran|0\ran^+$. We intend to use the primary to
highest weight null-state substitution $|j_0\ran\rar(J_0^-)
^{k+3}|j_1=\frac{k}{2}+1\ran$. Following our procedure we introduce
the parameter $\ep$ and take $j_1\rar j_1+\ep$ and $\ti{j}_0
\rar\ti{j}_0-\ep$ in order to preserve the zero eigenvalue of $J_0
^{3,tot}$. Acting on this state with $\hat{Q}$ will, in the
limit $\ep\rar0$, give us the non-trivial BRST invariant state  $c_0^-(J_0^-)
^{k+2}|j_1\ran|\ti{j}_0\ran|0\ran^+$. For ghost number minus
one we use the homotopy operator on the highest weight null-state to find
$b_0^-\sum_
{i=0}^{k+2}(-1)^i(J_0^-)^{k+2-i}(\jt_0^-)^i|j_1\ran|\ti{j}_0
\ran|0\ran^+$ in the cohomology.

In the second step we substitute $|j_1=\frac{k}{2}+1\ran\rar
J^+_{-1}|j_2=\frac{k}{2}\ran$ in the BRST invariant state at ghost number one.
After the usual steps of applying $\hat{Q}$ and taking
$\epsilon$ to zero, we find the resulting state $c_0^-c_{-1}^
-(J_0^-)^k|j_2\ran|\ti{j}_0\ran|0\ran^+$. We have here dropped
several terms which in the limit $\ep\rar0$ became null-states.
By inspection it is obvious
in this example that there exists only one state at ghost number two
for these representations, and hence the cohomology is here one-dimensional.

For ghost number minus two we will only give the null-state
$|\bar{N}_{-1}\ran$ of ghost number minus one, which is BRST exact
and from which one may construct
the state at ghost number minus two. Following our general
discussion above, we will use states at ghost number minus one,
and rewrite them as null-states. To construct the former, we
introduce the null-states
 $n_1|j_1\ran\equiv (J^+_{-1}(J_0^-)^2-(k+3)J^3_{-1}J_0^--\frac{(k+2)(k+
3)}{2}J_{-1}^-)|j_1=-\frac{k}{2}-1\ran$, $n_2|j_2\ran\equiv (J_0^-)
^{k+1}|j_2=\frac{k}{2}\ran$, $n_3|j_3\ran\equiv (J_0^-)^{k+3}|j_3=
\frac{k}{2}+1\ran$, $n_4|j_2\ran\equiv J^+_{-1}|j_2\ran$. In the
auxiliary sector we get $\ti{n}_1|\ti{j}_1\ran$ and $\ti{n}_4|
\ti{j}_4\ran$ from $n_1|j_1\ran$ and $n_4|j_4\ran$ by the
substitution $J\rar\jt$ and $k\rar-k-4$. Also $\ti{n}_2|\ti{j}_2\ran$
and $\ti{n}_3|\ti{j}_1\ran$ follow from $n_2|j_2\ran$ and $n_3|j_3\ran$
by substitution $J\rar\jt$. The relevant states
$s_1|j_1\ran|\ti{j}_1\ran|0\ran^+,s_3|j_3\ran|\ti{j}_1\ran|0\ran^+,
\ti{s}_2|j_2\ran|\ti{j}_2\ran|0\ran^+, \ti{s}_4|j_2\ran|\ti{j}_4 \ran|0\ran^+$
 for ghost number minus one may be found, using the homotopy
operator, from the null-states satisfying the equations
$\hat{Q}s_1|j_1\ran|\ti{j}_1
\ran|0\ran^+=(n_1+\ti{n}_1)|j_1\ran
|\ti{j}_1\ran|0\ran^+$, $\hat{Q}s_3|j_3\ran|\ti{j}_1\ran|0\ran^+=
(n_3+(-1)^{k+2}\ti{n}_3)|j_3\ran|\ti{j}_1\ran|0\ran^+$,
$\hat{Q}\ti{s}_2|j_2\ran|\ti{j}_2\ran|0\ran^+=(n_2+(-1)^k\ti{n}_2)
|j_2\ran|\ti{j}_2\ran|0\ran^+$, and $\hat{Q}\ti{s}_4|j_2\ran$ $
|\ti{j}_4\ran |0\ran^+=(n_4+\ti{n}_4)|j_2\ran|\ti{j}_4\ran|0\ran^+$.
We now use primary to null-state substitutions $|j_1\ran\rar
n_2|j_2\ran$, $|j_3\ran\rar n_4|j_2\ran$, $|\ti{j}_2\ran\rar
\ti{n}_1|\ti{j}_1\ran$, and $|\ti{j}_4\ran\rar \ti{n}_3
|\ti{j}_1\ran$, and demand that $\hat{Q}|\bar{N}_{-1}\ran
=\hat{Q}(a_1s_1n_2+a_3s_3n_4+
a_2\ti{s}_2\ti{n}_1+a_4\ti{s}_4\ti{n}_3)|j_2\ran
|\ti{j}_1\ran|0\ran^+=0$, where $a_i$ are constants. This is satisfied for
$a_3=-a_1$,
$a_2=-a_1$, and $a_4=(-1)^{k+2}a_1$. We have here used
$n_1n_2=n_3n_4$, and $\ti{n}_2\ti{n}_1=\ti{n}_4\ti{n}_3$, which
may be calculated explicitly or understood from the uniqueness
of highest weight null-states. All that remains is to use the
homotopy operator to construct $|\bar{S}_{-2}\ran$ from the
equation $|\bar{N}_{-1}\ran=\hat{Q}|\bar{S}_{-2}\ran$.

\end{document}